\documentclass{PoS}

\usepackage{amsmath}

\newcommand{\eq}{\begin{equation}}
\newcommand{\eqx}{\end{equation}}
\newcommand{\eqn}{\begin{eqnarray}}
\newcommand{\eqnx}{\end{eqnarray}}
\newcommand{\alg}{\begin{align}}
\newcommand{\algx}{\end{align}}

\newcommand{\f}[2]{\frac{#1}{#2}}

\newcommand{\lm}{\lambda}

\renewcommand{\th}{\theta}
\newcommand{\sg}{\sigma}

\newcommand{\nn}{{\cal N}}
\newcommand{\tr}{\mbox{\rm tr}\,}

\newcommand{\OO}[1]{{\cal O}\left(#1\right)}

\title{Non-perturbative Field Theory}

\ShortTitle{Non-perturbative Field Theory}

\author{\speaker{Romuald A. JANIK}%
\\
        Jagiellonian University\\
        E-mail: \email{janik@th.if.uj.edu.pl}}

\abstract{In this talk we review recent developments which enable us to use
techniques of integrable two-dimensional quantum field theories to solve exactly four dimensional N=4 gauge theory through the use of the AdS/CFT correspondence. By `solve' we mean here to find all anomalous dimensions of all operators of the theory for any value of the gauge coupling.
We illustrate the methods with the case of the Konishi operator, the simplest operator not protected by supersymmetry, whose perturbative anomalous dimension can be computed from the string side of the AdS/CFT correspondence and is in exact agreement with a direct gauge theory perturbative computation.}

\FullConference{European Physical Society Europhysics Conference on High
Energy Physics\\
                 July 16-22, 2009\\
                 Krakow, Poland}

\begin{document}

\section{Introduction}

Interacting quantum field theories form the basis of our description of the physical world, yet very often the simplicity of their definition or formulation is accompanied by very serious difficulties in extracting consequences. This happens in particular for QCD, where apart from a well understood perturbative regime there are lots of key questions which cannot be addressed using perturbative methods and which neccesitate analysis at strong coupling on a nonperturbative level. Various methods have been developed in the past for dealing with these problems including effective field theories, models of the QCD vaccuum or a direct {\it ab-initio} formulation using numerical lattice methods. Despite successes, however, still numerous open questions remain.

For gauge theories with some amount of supersymmetry, it is possible to obtain certain exact results for some special class of observables. Unfortunately, for generic observables this was not possible so far. This is in contrast to a wide class of interacting quantum field theories in two dimensions were such generic questions may be answered -- like the energy spectrum of the theory in finite volume, exact scattering amplitudes etc. By solvability we mean not neccessarily an analytical solution but more generally a reformulation into a well posed \emph{finite} mathematical problem like a set of integral equations etc. However the methods used in those cases depend in a crucial way on the two-dimensionality and do not extend to a higher number of dimensions.

Around 10 years ago a fascinating new approach to the study of nonperturbative properties of (mostly supersymmetric) gauge theories was discovered. The AdS/CFT correspondence \cite{adscft} identifies the maximally supersymmetric $\nn=4$ gauge theory with superstring theory in ten-dimensional curved $AdS_5 \times S^5$ spacetime. Its utility for studying nonperturbative physics lies in the fact that at strong gauge theory coupling, the dual string side becomes tractable. Commonly the AdS/CFT correspondence is understood as a useful tool only at strong coupling, however recent developments give hope that it may be effectively used for \emph{any} coupling leading to a full exact solution (at least for the spectrum) of the four dimensional $\nn=4$ gauge theory. Such knowledge would allow to trace the behaviour of various quantities all the way from strong to weak coupling making contact with conventional perturbative computations in gauge theory. The main ingredient which makes possible such an exact solution is the fact that the AdS/CFT correpsondence relates the physics of a 
\emph{four-dimensional} gauge theory to the \emph{two-dimensional} worldsheet quantum field theory of the superstring in $AdS_5 \times S^5$. In this talk we would like to review, in a very introductory way, some of these recent developments.  

\section{Integrability --- exactly solvable quantum field theories}

Basically the only quantum field theories which have been solved exactly are defined on a two-dimensional spacetime. These belong to two main classes: conformal field theories (CFT) and integrable field theories. For many examples of the first class we have fairly complete information including both the spectrum and correlation functions. However these theories do not have a mass scale, and in order to consider solvable quantum field theories which are massive one is led to the second class mentioned above -- integrable field theories. Usually we do not have as complete information as for CFT's, in particular regardng correlation functions, however other observables may be obtained exactly -- in particular the S-matrix for scattering of asymptotic states, and, for many theories, the spectrum of energy levels for the theory defined on a finite size cylinder.

The quintessential example is Sine-Gordon theory defined by the lagrangian
\eq
S=\int d^{{2}}x \left\{  \f{1}{2} \partial_\mu \Phi \partial^\mu \Phi +
\f{\mu^2}{\beta^2} \cos(\beta \Phi) 
\right\}
\eqx
We see that expanding $\cos(\beta \Phi)$ for small $\beta$ generates an infinite set of vertices for $\Phi$, so one can develop perturbation theory. On the other hand, for strong coupling, one can exhibit classical solitons and  antisolitons, analyze their semiclassical quantization etc.

However it turns out that the exact S-matrix for the scattering of solitons (and anti-solitons and their possible bound states) can be found exactly and \emph{analytically}. E.g. the soliton-soliton S-matrix is given by the explicit formula
\eq
S(\th)=-\exp \left\{ 2\int_{0}^\infty \f{dk}{k} \sin\left(k\th \right) 
\f{\sinh (p-1)k}{2\cosh k \sinh p k}  \right\}
\eqx
where
\eq
p=\f{\beta^2}{8\pi -\beta^2}
\eqx  
The above expression has a very nontrivial expansion both at weak and strong coupling and it would be inconcievable to determine it by resumming results from either end. The reason why obtaining such explicit formulas is possible is the property of integrability of the Sine-Gordon theory. 

Integrability in two-dimensional QFT means the existence of a sufficient number of higher spin (or non-local) conserved charges. Their existence severly constraints the dynamics of the theory implying in particular that the number of particles in a scattering process is conserved (no particle production). Moreover their momenta are only permuted, and most importantly the scattering of $N$ particles factorizes into products of $2\to2$ scattering. Hence the whole dynamical information on the scattering amplitudes lies in the two particle S-matrix, which must moreover satisfy a nonlinear functional equation - the so-called Yang-Baxter Equation (YBE). This equation, together with a very basic input on the symmetries of the theory and crossing and unitarity allows to find the exact, fully nonpertutbative S-matrix, very often in an analytical form \cite{ZamZam}.

Unfortunately, if one would try to impose similar additional conservation laws on a quantum field theory defined in a higher number of dimensions one would obtain a trivial (free) theory. From this point of view it would seem that the two dimensional integrable quantum field theories remain just a very interesting but nevertheless theoretical curiosity. So it seemed that the methods used for extracting nonperturbative information in such theories would not work in higher dimensions.

However the AdS/CFT correspondence turns out to provide an avenue for the application of two-dimensional quantum field theories for four-dimensional physics.

\section{The AdS/CFT correspondence and $\nn=4$ SYM}

The AdS/CFT correspondence introduced in \cite{adscft}, states the exact equivalence between $\nn=4$ supersymmetric gauge theory and superstring theory in an 
$AdS_5 \times S^5$ background. The correspondence is fascinating for a variety of reasons. Firstly, on a purely theoretical ground, it relates two completely different theoretical constructions -- a four-dimensional gauge theory and a \emph{string theory} living in a ten dimensional background. Secondly, the relation between the parameters of the two theories is such that strong coupling gauge theory physics gets mapped to (semi-)classical strings/gravity, and so one can use these methods to extract nonperturbative information about 
the $\nn=4$ gauge theory.

In the more than 10 years since the discovery of AdS/CFT, these methods have been used for a wide variety of physics problems. Some notable examples are applications to the study of hot gauge theory matter, which has implications for our understanding of quark-gluon plasma produced in heavy-ion collisions \cite{QGP}, and a very recent research program which aims at understanding strongly coupled three dimensional theories exhibiting similar properties as some condensed matter systems \cite{CMT}.

Let us mention that the very reason for which the AdS/CFT correspondence has been so useful in transforming very hard nonperturbative problems into tractable questions on the dual side makes it very difficult even to test quantitatively let alone prove. The calculations which can be carried out on the gauge theory side using perturbation theory correspond to highly quantum stringy regime on the AdS side. Recently, however, there was a lot of progress in this respect which we will review in the present talk.

Let us briefly analyze the picture emerging from the AdS/CFT correspondence on gauge theory physics at strong coupling. We will use the duality to isolate relevant degrees of freedom at strong coupling. Since the duality postulated by AdS/CFT involves string theory, we have to deal with strings. Recall that vibrational modes of the string in $AdS_5 \times S^5$ correspond to particles/fields in $AdS_5 \times S^5$. A closed string always has a massless sector consisting of gravitons and other fields of type IIB supergravity. In addition it has a whole (infinite) tower of massive fields. When the coupling constant of the $\nn=4$ gauge theory is large, those massive fields become very heavy and for most physical proccesses of interest decouple from the massless supergravity modes. Hence at strong coupling, for the majority of problems, the relevant degrees of freedom are gravitational and many gauge theory problems get mapped to General Relativity `geometrical' questions. If we decrease the coupling, massive string modes beome important and then the full stringy dynamics has to be taken into account. Unfortunately this is much less understood in general.  Incidentally let us note that in the case of QCD, if a string dual ala AdS/CFT exists, presumably there will never be a clear separation into gravity and massive string modes thus making the dual picture much more complex and difficult to treat. In the rest of this talk, we will always stick to the case of the
$\nn=4$ supersymmetric gauge theory, but we will go to the regime where string modes will be important.

The complexity described above which emerges at intermediate coupling does not seem, at first glance, to give a hope for an exact treatment at any coupling. To see that fortunately this is not the case let us return to the original form of the AdS/CFT correspondence equating the supersymmtric gauge theory with the theory of \emph{strings} in $AdS_5 \times S^5$.

\section{Towards an exact solution for any coupling}

Consider a single closed string in $AdS_5 \times S^5$ spacetime. The worldsheet of the closed string has the topology of a cylinder and it can be parametrized by $(\tau,\sg)$ coordinates. Then the embedding in $AdS_5 \times S^5$ can be defined by giving the embedding coordinates as functions of the cylinder coordinates $X^\mu(\tau,\sg)$. From the point of view of the cylinder, these are just (scalar) fields on the cylinder. These are then promoted to 
\emph{quantum fields} on a cylinder with an action determined by the geometry of the background $AdS_5 \times S^5$ spacetime. In reality, for the superstring in $AdS_5 \times S^5$, one has to include fermionic fields, perform a gauge fixing of local symmetries etc. At the end of the day superstring theory in $AdS_5 \times S^5$ amounts to\footnote{Here we neglect interaction between strings, which can be justified for most problems when we take the 
$N_c \to \infty$ limit on the gauge theory side.} a specific \emph{two dimensional quantum field theory} defined on a cylinder.

It is at this stage that we see the emerging connection between four dimensional gauge theory and two dimensional quantum field theories. Moreover, the specific two dimensional worldsheet QFT of the superstring in $AdS_5 \times S^5$ turns out to be integrable (which was first established at the classical level in \cite{BPR}) thus opening up the fascinating possibility of using methods of 2D integrable QFT's in order to solve exactly a 4D gauge theory for any value of the gauge theory coupling constant. On the weak coupling side of the $\nn=4$ gauge theory there have been independent discoveries of integrability \cite{MZ}. This gives reassurance that the 2D worldsheet QFT is integrable also at the fully quantum level.

Of course one has to qualify what do we mean by `solve'. 
Since $\nn=4$ gauge theory is a conformal field theory, the most natural question to ask is what are the eigenvalues of the dilatation operator or equivalently, what are the anomalous dimensions of \emph{all} operators for \emph{any} coupling $\lm \equiv g^2_{YM}N_c$. Naturally, on the gauge theory side this is a highly nontrivial problem. The same question when expressed in terms of the corresponding string theory is to find the energy levels of a string in $AdS_5 \times S^5$, or equivalently, to find the energy levels of the corresponding worldsheet QFT on a cylinder of any fixed size\footnote{The (integer!) cylinder size in the light cone gauge fixed string corresponds to a specific charge of the corresponding gauge theory operator.}. These two equivalent questions are one of the main goals of the integrability program in AdS/CFT.

Let us recall that the first step toward solving a two-dimensional integrable quantum field theory is its solution on a plane (at infinite volume) i.e. the identification of possible (massive) asymptotic states and the determination of the S-matrix for the scattering of these states. This part of the program has been currently carried out and we believe that the spectrum of asymptotic states of the 2D worldsheet QFT involves an infinite set of particles labelled by $Q=1\ldots \infty$ \cite{DoreyBound}, and the S-matrix of their scattering is known exactly for any value of the coupling $\lm$ 
\cite{S,BS,B,CROSS,BHL,BES,BoundS}.

Let us emphasize that this part of the program has been highly nontrivial and \emph{not} a straightforward generalization of the well known relativistic case, as the 2D worldsheet QFT has some exceptional and unique properties -- in particular 2D Lorentz invariance is broken, the 2D energy and momentum are intertwined with the global symmetry algebra \cite{B}, a torus arises as the parameter space of rapidities \cite{CROSS}, the S-matrix does not have difference property \cite{S,BS,B}. However all those difficulties have been overcome, but progress required multiple insights both from perturbative gauge theory \cite{oneloop1,oneloop2,BDS} as well as from classical string solutions and their integrable structures 
\cite{KMMZ,AFS} etc. 

Unfortunately, even an exact solution of the 2D worldsheet QFT on the plane is not enough for the application to AdS/CFT since there we are dealing with a closed string which means that the 2D QFT is defined not on a plane but on a cylinder of finite size. Let us now analyze briefly how the energy levels of the cylinder theory may be found from the knowledge of the infinite volume solution.

\subsection*{Large volume energy spectrum}

When the cylinder is very large, a very natural universal physical picture emerges for the energy levels of the theory. Consider an $N$-particle state on a cylinder of a very large size $L$. We may treat the particles as very well separated with given momenta $p_i$ and consider their quantum mechanical wavefunction. Now we take one of the particles and move it around the circumference of the cylinder. The wavefunction will get a phase shift 
$e^{i p_i L}$ from the translation and additional phase factors from scattering of this particle with the remaining ones. Since we are in an integrable theory these phase factors factorize and one is led to Bethe equations
\eq
\label{e.bethe}
e^{i p_i L} =\prod_{k \neq i} S(p_k,p_i)
\eqx
In reality, for nondiagonal scattering one has to be more careful and one gets a system of coupled equations. The Bethe equations which arise for the worldsheet 2D QFT are exactly the Asymptotic Bethe Ansatz of Beisert and Staudacher \cite{BS}. Now one has to solve \eqref{e.bethe} for the momenta $p_i$ and find the energy from
\eq
E=\sum_{i=1}^N E(p_i)
\eqx

One important point has to be made here -- Bethe equations of an identical form (but with various different S-matrices) arise also in the context of spin chains where they give the \emph{exact} energy levels for \emph{any} size $L$. In the context of a 2D quantum field theories however, these equations are just a first approximation. In a QFT one cannot use a purely quantum mechanical picture and there will be important modifications due to virtual corrections. 
Thus the long range spin chain corresponding to the Asymptotic Bethe Ansatz can be understood just as the leading large volume limit of the worldsheet 2D QFT. 

\subsection*{First corrections}

The first corrections to the energy levels coming from Bethe Ansatz are called 
L{\"u}scher corrections, who derived the first corrections to single particle states for relativistic quantum field theories \cite{Luscher}. These corrections are completely expressed in terms of the infinite volume data of the theory -- the S-matrix and the spectrum of asymptotic states. Recently, these corrections were generalized to the nonrelativistic case of the AdS string \cite{JL}, as well as to multiparticle states \cite{Konishi4}. They typically arise due to a virtual particle circulating around the circumference of the cylinder.
Interestingly, deviations from the Bethe ansatz have a very natural interpretation on the gauge theory side. There, the calculation of anomalous dimensions through two point functions of single trace operators leads to graphs with the topology of a cylinder in the planar limit. The deviations from the Bethe Ansatz correspond to graphs with propagators encircling the whole cylinder. This is reminiscent of a L\"uscher graph like the one shown in fig.~1 below. The loop order at which these corrections arise can be quantitatively understood both from the gauge theory and from the 2D worldsheet QFT perspective of the string in $AdS_5 \times S^5$ \cite{AJK}.

\subsection*{Exact spectrum}

If one decreases the size of the cylinder even more, one would have to deal with multiple virtual corrections. Luckily for many relativistic integrable field theories one can resum \emph{all} corrections through the formalism of Y-systems, Thermodynamic Bethe Ansatz (TBA) or Nonlinear Integral Equations (NLIE), all of which lead generically to a system of (perhaps infinite) coupled nonlinear integral equations. Knowing the solution, say numerically, enables one to obtain the energy level to any desired accuracy. 

The relevant formulation for the AdS case have been recently proposed in 
\cite{exact1,exact2,exact3}. Currently these systems are quite complicated and involve an infinite set of equations and unknowns. Much of the current research aims at reducing the complexity to something more managable.

\begin{figure}[t]
\begin{center}
 \includegraphics[height=4cm]{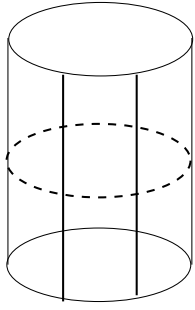}
\end{center}
\caption{The leading L\"uscher graph contributing to the Konishi operator at four loops. The dashed line represents all asymptotic states of the theory while the two vertical lines correspond to the two particles forming the Konishi state in the two dimensional string worldsheet QFT.}
\end{figure}

\section{A case study: the Konishi operator}

In this last part of the talk, we would like to exhibit the progress made along these lines for the simplest operator in $\nn=4$ which is not protected by supersymmetry -- the Konishi operator. The reason for considering this operator is that much effort has been invested in direct perturbative computations of its anomalous dimension up to four loops - the order where first corrections to the Bethe ansatz appear for the first time.

The Konishi operator is the operator $\tr \Phi_i^2$, where $\Phi_i$ are the adjoint scalars of $\nn=4$ SYM. It is more convenient to consider the operator 
$\tr DZDZ- \tr ZD^2Z$ with exactly the same anomalous dimension. Here $D$ is a covariant derivative along a light-cone direction and $Z$ is a complex scalar, say $Z=\Phi_1+i\Phi_2$. On the string side, this corresponds to a two particle state (two $D$'s) on a cylinder of size 2 (two $Z$'s). A solution of Bethe ansatz leads to 
\eq
\Delta_{Bethe}=4+12g^2-48g^4+336 g^6-(2820+288\zeta(3))g^8+\ldots
\eqx
where $g^2 \equiv \lm/(16\pi^2)$ with $\lm=g^2_{YM} N_c$ being the `t Hooft coupling of the gauge theory. The coefficients of $g^2$, $g^4$ and $g^6$ agree with perturbative computations on the gauge theory side at 1-, 2- and 3-loop order. The first correction to $\Delta_{Bethe}$ due to L\"uscher corrections will appear first at order $g^8$ (four loops). This correction can be seen to follow from a single `L\"uscher graph' for the two-dimensional worldsheet QFT shown in fig.~1 \cite{Konishi4}. One has to sum over all asymptotic states of the theory (summation over $Q$ below), integrate over their momenta $q$ and use the exact S-matrix between the virtual particles and the two particles forming the Konishi state. One gets
\eq
\Delta^{(4-loop)}_{wrapping}=-\sum_{Q=1}^{\infty}\int_{-\infty}^{\infty}\!
\frac{dq}{2\pi} Y_Q(q)
\eqx
where $Y_Q(q)$ is a relatively simple rational function of $q$ and $Q$.
This can be evaluated by residues giving
\eq
\label{e.sumQ}
\Delta^{(4-loop)}_{wrapping}= \sum_{Q=1}^{\infty} \left\{ -\frac{num(Q)}{\left(9 Q^4-3
   Q^2+1\right)^4 \left(27 Q^6-27 Q^4+36 Q^2+16\right)}
+\frac{864}{Q^3}-\frac{1440}{Q^5} \right\}
\eqx
where the numerator $num(Q)$ is given by
\begin{align}
num(Q)=& 7776 Q (19683 Q^{18}-78732 Q^{16}+150903
   Q^{14}-134865 Q^{12}+ \nonumber\\
&+1458 Q^{10}+48357 Q^8-13311
   Q^6-1053 Q^4+369 Q^2-10)
\end{align}
The above may be summed giving 
\eq
\Delta^{(4-loop)}_{wrapping}=(324+864\zeta(3)-1440 \zeta(5))g^8
\eqx
This exactly agrees with a direct perturbative computation of the four loop Konishi wrapping effects using supergraphs in \cite{FSSZ} (comprising around 120 supergraphs). This was confirmed by a perturbative computation using Feynman graphs in components \cite{Velizhanin08} (including 131015 graphs).

Let us emphasize that the two computations, one using L\"uscher graphs in the two dimensional worldsheet QFT of the string, and the other one, using Feynman graphs/supergraphs in the four dimensional $\nn=4$ gauge theory are very complex and have apparently nothing in common. Yet they give an exact agreement. This is a very nontrivial quantitative test of the AdS/CFT correspondence. In addition this tests the correctness of our knowledge of the ininite volume solution of the 2D QFT, and it also shows that the Bethe ansatz for the gauge theory is violated in just the way expected from a 2D quantum field theoretical perspective.

Currently the analytical L\"uscher computation from string theory in $AdS_5 \times S^5$ has been extended to 5 loop orders \cite{Konishi5} giving the Konishi anomalous dimension to this order to be 
\eqn
\Delta&=&\Delta_{Bethe}+\Delta_{wrapping}=4+12\,g^2-48\,g^4+336\, g^6+ 96 (-26 + 6 \,\zeta(3) - 15 \,\zeta(5))\, g^8 
 \\ \nonumber
&& -96 (-158 - 72 \,\zeta(3) + 54 \,\zeta(3)^2 + 90 \,\zeta(5) - 315\, \zeta(7))\, g^{10} + \OO{g^{12}}
\eqnx 

Very recently the proposed exact formulation of the AdS worldsheet QFT in terms of the Y-system has been used to propose a set of nonlinear integral equations leading to the exact energy of the Konishi state for any value of the coupling constant. The result of the numerical solution of \cite{KonishiNum} are shown in fig.~2.

\begin{figure}[t]
\begin{center}
 \includegraphics[height=4cm]{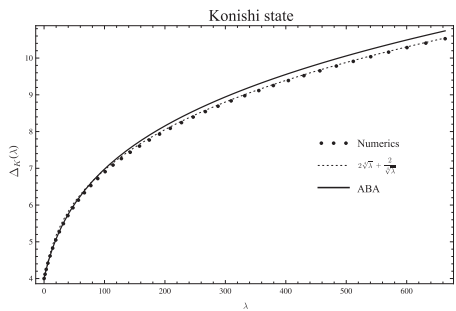}
\end{center}
\caption{The results of the numerical evaluation of the anomalous dimension of the Konishi operator from \cite{KonishiNum}.}
\end{figure}

\section{Outlook}

In this talk we described recent developments which, through the AdS/CFT correspondence, enable us to use the techniques of exactly solvable integrable two-dimensional quantum field theories to move very close to an exact solution (for the spectrum of anomalous dimensions for all operators for any value of the coupling constant) of the \emph{four-dimensional} $\nn=4$ gauge theory.
Currently it seems that this is definitely within reach with proposed systems of integral equations in some sectors of the theory. Very probably this may be simplified, leading to new analytical insights, and presumably the structure of most general source terms corresponding to competely generic operators will be worked out in the near future.

The techniques of integrability allow to perform computations on the string side of the AdS/CFT correspondence which are valid at weak gauge theory coupling. Amazingly these computations in the 2D worldsheet QFT are much simpler than direct perturbative computations in the gauge theory. This may signify that there is some structure in the perturbative expansion of gauge theory which has not been understood so far. It is quite amusing that one can recover pertubative results in gauge theory using nonperturbative methods in the 2D worldsheet QFT. In addition, these methods will be applicable for any coupling all the way to the strongly coupled nonperturbative regime of the 
$\nn=4$ gauge theory. 

Finally let us note that $\nn=4$ gauge theory has been dubbed {\it `the harmonic oscillator of four dimensional gauge theories'}. If we can solve it exactly we may hope to use this as a point of departure in our understanding of other gauge theories with QCD being the ultimate goal.

\bigskip

\acknowledgments

\noindent{}RJ was supported by Polish science funds during 2009-2011 as a research project (NN202 105136) and Marie Curie ToK KraGeoMP (SPB 189/6.PRUE/2007/7).

\end{document}